\documentclass{aipproc}
\usepackage{amssymb} 
\layoutstyle{8x11single}
\begin{document}

\title{Relativistic $r$-modes and shear viscosity}


\author{L. Gualtieri}{
  address={Dipartimento di Fisica ``G. Marconi, Universit\`a di Roma 
``La Sapienza'' \\ and Sezione INFN ROMA1, piazzale A. Moro 2, 00185
Roma, Italy}}

\author{J.A. Pons}{
  address={Departament de F\'isica Aplicada, Universitat d'Alacant, 
Apartat de correus 99, 03080 Alacant, Spain}}

\author{J.A. Miralles}{
  address={Departament de F\'isica Aplicada, Universitat d'Alacant, 
Apartat de correus 99, 03080 Alacant, Spain}}

\author{V. Ferrari}{
  address={Dipartimento di Fisica ``G. Marconi, Universit\`a di Roma 
``La Sapienza'' \\ and Sezione INFN ROMA1, piazzale A. Moro 2, 00185
Roma, Italy}}

\begin{abstract}
We derive the relativistic equations for stellar perturbations,
including in a consistent way shear viscosity in the stress-energy
tensor, and we numerically integrate our equations in the case of
large viscosity. We consider the slow rotation approximation, and we
neglect the coupling between polar and axial perturbations. In our
approach, the frequency and damping time of the emitted gravitational
radiation are directly obtained. We find that, approaching the
inviscid limit from the finite viscosity case, the continuous spectrum
is regularized. Constant density stars, polytropic stars, and stars
with realistic equations of state are considered.  In the case of
constant density stars and polytropic stars, our results for the
viscous damping times agree, within a factor two, with the usual
estimates obtained by using the eigenfunctions of the inviscid
limit. For realistic neutron stars, our numerical results give viscous
damping times with the same dependence on mass and radius as
previously estimated, but systematically larger of about 60\%.
\end{abstract}

\maketitle


\section{Introduction}
Rotating neutron stars are a promising source of gravitational waves.
Indeed, their oscillations can become unstable, converting their
rotational energy into gravitational radiation.  It is of particular
interest the instability of the $r$-modes, which are oscillation modes
of the star associated with the Coriolis force, and can become
unstable at any rotation rate of the star \cite{And98}.

It has been suggested that $r$-mode instability could produce waves
strong enough to be detected by LIGO and VIRGO\cite{detect}.
Furthermore, these instabilities are thought to play an important role
in astrophysics. For instance, they may be relevant in spinning-down
newly born neutron stars (and quark stars, if they actually exist)
\cite{AKS,LOM}, and could be responsible for preventing the spin-up of
accreting neutron stars in low mass $X$-ray binaries \cite{BIL,WAG}.

In spite of the extensive literature dedicated to the study of the
problem, our present understanding of $r$-mode instability of
compact stars is still incomplete, with many controversial issues.
  
It has been pointed out \cite{Kojimar,RK1,RK2} that, working at first
order in perturbation theory and neglecting the coupling between
different harmonics, leads to the existence of a continuous spectrum
and makes doubtful the existence of the modes. However,
non-perturbative numerical simulations \cite{Nick,loic} seem to
indicate that $r$-modes do exist, so that the continuous spectrum may
be interpreted as an artifact due to an inconsistency of the
perturbative expansion when $\sigma\sim\Omega$.  With this motivation,
more sophisticated methods to solve the eigenvalue problem have been
developed \cite{LAF1,LAF2,LAW}.

Another interesting issue, which up to now has not been fully
understood, is the role of viscosity in $r$-mode oscillations.  It is
believed that bulk/shear viscosity limit the instability at high/low
temperatures, respectively. But a complete understanding of this
mechanism is still missing. Furthermore, it would be interesting to
understand if the inclusion of viscosity does affect the existence of
the continuous spectrum, as suggested for example in \cite{LAW}.

Here we report the results of \cite{us}, where we have studied the
effect of introducing shear viscosity to the $r$-mode oscillations
frequencies and damping times.  We have introduced shear viscosity
from the beginning in the stress-energy tensor, solving 
perturbatively the Einstein's equations; in this way we have been able
to evaluate consistently both the frequency and the damping time of
the mode, for the first time . A similar self-consistent inclusion of
the heat transfer corrections has recently been done in \cite{GPM04}.
  
Furthermore, we have found that by including a small amount of
viscosity we are able to regularize the continuous spectrum.  
\section{Formulation of the problem}   
  
We have considered a star rotating uniformly with angular velocity $\Omega$. 
At first order in $\Omega$ (or, more precisely, at first order in the 
rotational parameter $\epsilon\equiv\Omega/\Omega_N$ with 
$\Omega_N\equiv\sqrt{2M/R^3}$), the stationary background is described 
by the metric \cite{Hartle,HartleThorne}
\begin{equation} 
ds^2=g^{(0)}_{\mu\nu}dx^\mu dx^\nu= 
-e^{\nu(r)}dt^2+e^{\lambda(r)}dr^2+r^2(d\vartheta^2+\sin^2\vartheta 
d\varphi^2) -2r^2\omega(r)\sin^2\vartheta d\varphi dt\,,\label{metric} 
\end{equation}  
where $\omega(r)$ represents the dragging of the inertial frames.  It 
corresponds to the angular velocity of a local ZAMO (zero angular 
momentum observer), with respect to an observer at rest at infinity. 
The $4$--velocity of the fluid is simply 
$u^{(0)\mu}=(e^{-\nu/2},0,0,\Omega e^{-\nu/2})$, and the stress-energy 
tensor is 
\begin{equation}   
T^{(0)}_{\mu\nu}=(\rho+p)u^{(0)}_{\mu}u^{(0)}_{\nu}+pg^{(0)}_{\mu\nu}\,.  
\end{equation}  
We have assumed that viscosity does not affect the stationary
axisymmetric background, because the shear tensor do vanish
there. Therefore, in this regime the Einstein's equations reduce
to the standard TOV (Tolman-Oppenheimer-Volkoff) equations plus a
supplementary equation for the frame dragging $\omega(r)$:
\begin{equation}   
\bar\omega_{,rr}-\left(4\pi(\rho+p){e^\lambda} 
r-\frac{4}{r}\right)\bar\omega_{,r} 
-16\pi(\rho+p){e^\lambda}\bar\omega=0\label{hartleeq} 
\end{equation}  
where $\bar\omega(r)\equiv\Omega-\omega(r)$.  

Taking into account viscosity, the stress-energy tensor  
has the form (see for instance \cite{MTW}):
\begin{equation}   
T_{\mu\nu}=(\rho+p)u_{\mu}u_{\nu}+pg_{\mu\nu}-2\eta\sigma_{\mu\nu}  
-2\zeta g_{\mu\nu}u^{\alpha}_{~;\alpha}\,.  
\end{equation}  
Here $\eta,\,\zeta$ are the shear and bulk viscosity coefficients (see 
\cite{CL} for a discussion on the meaning of these 
coefficients), and 
\begin{equation}   
\sigma_{\mu\nu}\equiv\frac{1}{2}g^{\rho\lambda}\left(u_{\mu;\rho}  
P_{\nu\lambda}+u_{\nu;\rho}P_{\mu\lambda}\right)-\frac{1}{3}u^{\rho}_{~;\rho}  
P_{\mu\nu}  
\end{equation}  
with $P_{\mu\nu}$ being the projector onto the subspace orthogonal to 
$u_{\mu}$ 
\begin{equation}   
P_{\mu\nu}\equiv g_{\mu\nu}+u_\mu u_\nu\,.  
\end{equation}

We have focused our investigation to $r$-modes, which are
perturbations with axial symmetry. Therefore, we have considered axial
perturbations of the metric (\ref{metric}), expanded in tensor
spherical harmonics, and we have solved the Einstein's equations,
linearized around the background, for such perturbations:
\begin{equation}   
\delta G_{\mu\nu}=8\pi \delta T_{\mu\nu}\,.
\end{equation}
This approach allows to determine the complex frequency of the
quasi-normal modes of the star (and then, of the emitted gravitational
radiation)
\begin{equation}
\sigma=2\pi\nu+\frac{\rm i}{\tau}
\end{equation}
where $\nu$ is the real frequency of the mode, and $\tau$ is its
damping time.

We have neglected the coupling between perturbations with harmonic
indexes $l$ and $l\pm1$, as in \cite{RK2,Kojimapol}. Under this
approximation, bulk viscosity is not coupled to axial perturbations,
because it enters into the equations only through the axial-polar
$l\leftrightarrow l\pm 1$ couplings. Thus, we only have studied the
effects of shear viscosity, leaving the investigation of the effects
of bulk viscosity for future work.
  
We also have considered the Cowling and the Newtonian limits of our
equations, in order to make a detailed comparison with previous works.

Because of numerical problems, we cannot integrate our system of
equations for $\eta\lesssim10^{-6}$ km$^{-1}$. \footnote{Usually the
typical viscosity at $T=10^7K$ and $\rho=10^{15}$g/cm$^3$ is $\eta\sim
10^{23}~{\rm g/cm/s}=2.4\cdot10^{-11}$km$^{-1}$.}

\section{Results}   
\label{results}
Here we discuss the results of the numerical integration of our
perturbative equations, carried out in \cite{us}.  We have studied
different kinds of stellar models: constant density stars, polytropic
stars and realistic neutron stars. In each case, we have compared the
Newtonian limit, the Cowling approximation, and the relativistic
calculation, in order to establish the qualitative differences between
the three approaches and to compare to previous literature on the
subject.
\subsection{Homogeneous stars}  
  
In the case of constant density stars, it has been shown that the
frequencies of the $r-$modes lay outside the continuous spectrum (see
e.g. \cite{KS99} for the full GR results).  We consider a
uniform density star with a central energy density of $10^{15}$
g/cm$^3$, mass of 1.086 $M_\odot$ and radius of 8.02 km ($M/R=0.2$),
with the rotational parameter $\epsilon=0.3$. In Fig. 1 we show the
$r$--mode frequency versus the viscosity parameter $\eta$, assumed to
be constant throughout the star.  The Newtonian, Cowling, and General
Relativistic results are shown by dotted, dashed, and solid lines,
respectively.  The Newtonian and GR frequencies in the inviscid limit
(1064 Hz and 1144 Hz) are indicated as dotted and solid horizontal
lines, while the shadowed region indicates the continuous spectrum.
As shown in Fig. 1, as the viscosity decreases the inviscid Newtonian
and GR results are recovered, while the frequency in the Cowling
approximation falls inside the continuous spectrum.

Since the convergence to the inviscid limit is reached for $\eta
\approx 10^{-5}$ km$^{-1}$, the mode frequency in the Cowling
approximation can be found extrapolating the dashed line for $\eta
\rightarrow 0$, and the corresponding value is 1244 Hz, about a ten
percent larger than the GR value.  It should be stressed that the mode
frequency in the Cowling approximation in the inviscid limit had never
been possible before for perfect fluids.

We stress that, as shown in Fig. 2, the Newtonian, Cowling and GR
damping times coincide. All damping times differ for less than 20\%
and show a $1/\eta$ behaviour, as expected from previous estimates in
the literature. In particular, we can compare our results with the
analytic formula of \cite{CL}, resulting from Newtonian estimates of
the dissipative time scale of the shear viscosity
\begin{eqnarray}  
\tau = \frac{\rho R^2}{(l-1)(2l+1)\eta}.   
\label{tau0}  
\end{eqnarray}  
For our model $\tau = 3\times 10^{-8}/\eta$, with $\eta$ 
in km$^{-1}$ and $\tau$ in s.  This estimate is also shown in Fig. 2 
(thin solid line) and it is, surprisingly, in better agreement with 
the GR results than with the Newtonian ones.   
\begin{figure}  
\includegraphics[height=.3\textheight]{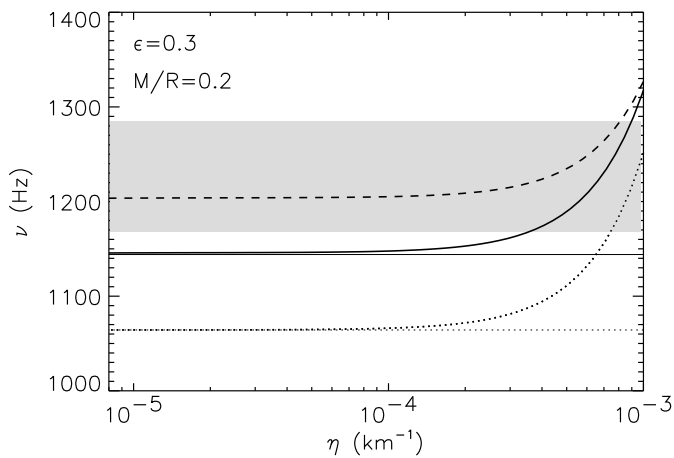}
\caption{The $r-$mode frequency as a function of the shear viscosity 
parameter $\eta$ for a uniform density model with $M/R=0.2$ and 
rotational parameter $\epsilon=0.3$.  The Newtonian, Cowling and GR 
results are shown, respectively, with dotted, dashed, and solid lines. 
The thin horizontal lines indicate the corresponding Newtonian and GR 
inviscid limits, while the continuous spectrum is the shadowed region.} 
\label{fig1a}  
\end{figure}  
\begin{figure}  
\includegraphics[height=.3\textheight]{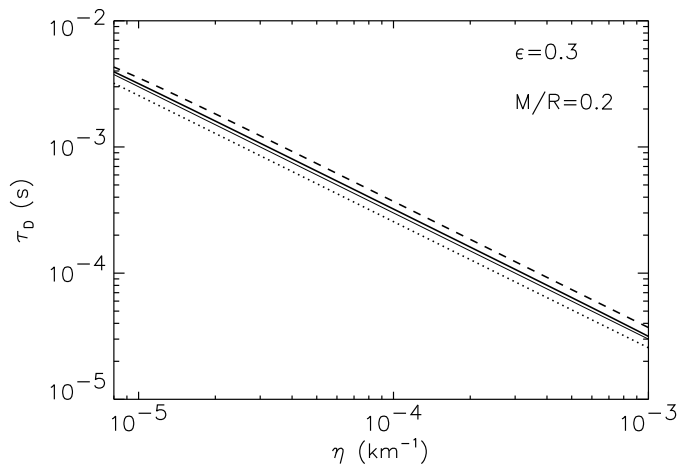}
\caption{The $r-$mode damping time as a function of the shear 
viscosity parameter $\eta$ for a uniform density model with $M/R=0.2$ 
and rotational parameter $\epsilon=0.3$.  The Newtonian, Cowling and 
GR results are shown, respectively, with dotted, dashed, and solid 
lines.  The thin solid line shows the simple estimate obtained using 
Eq. (\ref{tau0}). } 
\label{fig1b}  
\end{figure}  
\subsection{Polytropic Stars}  
\begin{figure}  
\includegraphics[height=.3\textheight]{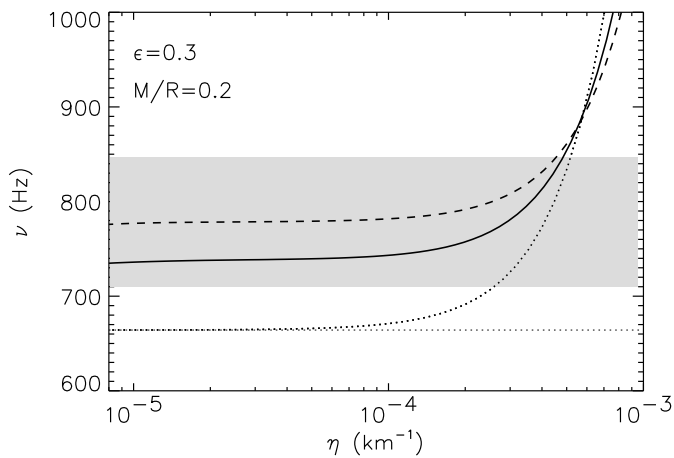}
\caption{The $r-$mode frequency as a function of the shear viscosity 
parameter $\eta$ for a polytropic star with $n=1$, $M/R=0.2$ and 
rotational parameter $\epsilon=0.3$.  The Newtonian, Cowling and GR 
results are shown, respectively, with dotted, dashed, and solid lines. 
The horizontal dotted line indicates the corresponding Newtonian 
inviscid limit, while the continuous spectrum is the shadowed region.} 
\label{fig2a}  
\end{figure}

\begin{figure}  
\includegraphics[height=.3\textheight]{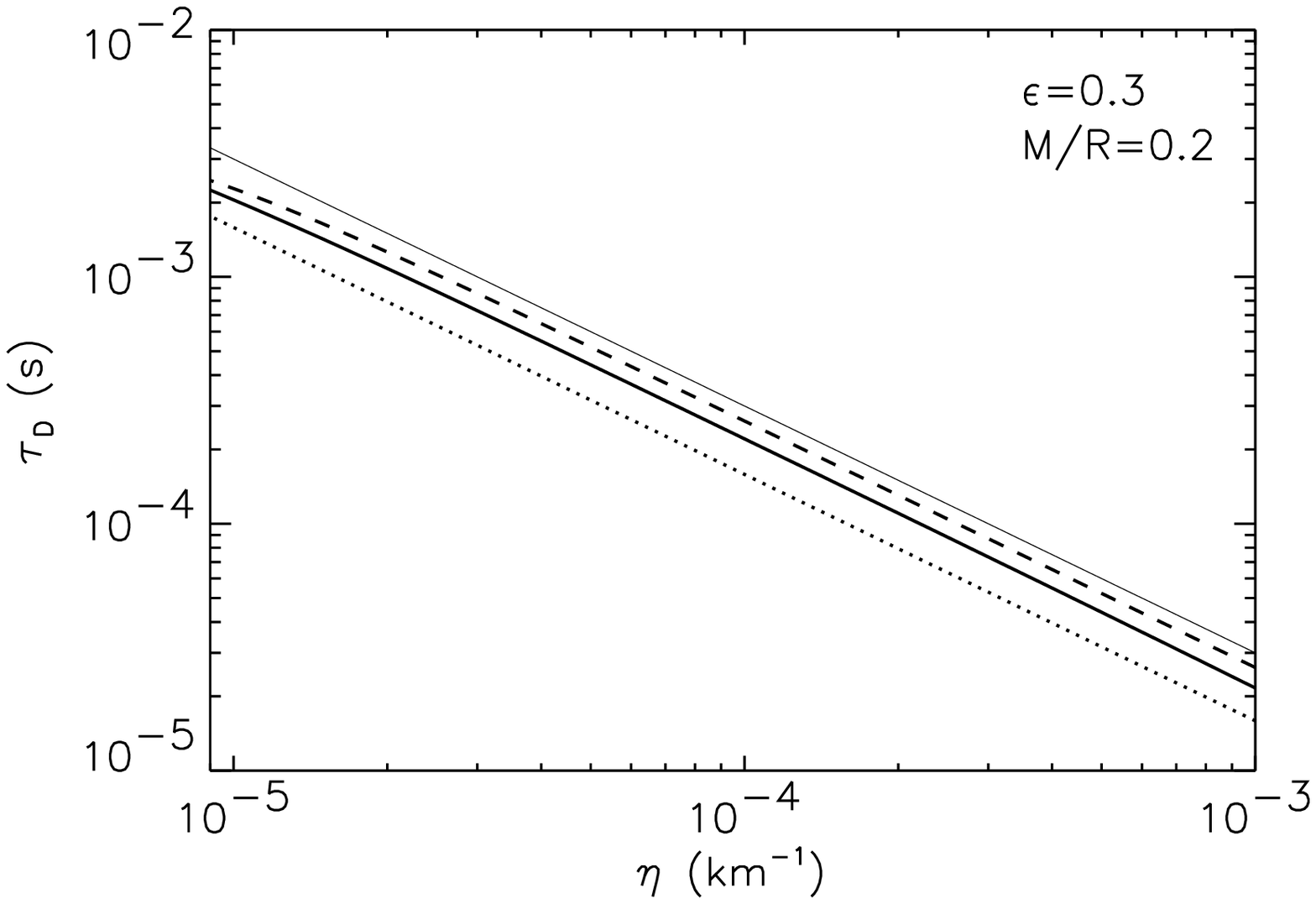}
\caption{The $r-$mode damping time as a function of the shear 
viscosity parameter $\eta$ for a polytropic star with $n=1$, $M/R=0.2$ 
and rotational parameter $\epsilon=0.3$.  The Newtonian, Cowling and 
GR results are shown, respectively, with dotted, dashed, and solid 
lines.  The thin solid line shows the simple estimate obtained using 
Eq. (\ref{tau0}). } 
\label{fig2b}  
\end{figure}

Here we consider polytropic stars. For these models the $r-$mode was
found to disappear for $n\ge 0.8$ \cite{RK1,RK2}.  Only for very
compact stars $n<0.8$, the $r-$mode frequency lies outside the
continuous spectrum and could be found. We have considered a
polytropic model with $n=1$, and the same compactness and rotation
parameter as in previous section ($M/R=0.2, \epsilon=0.3$). This model
has mass $M=1.74 M_\odot$ and $R=12.86$ km. In the inviscid case we do
not find the $r-$mode because it lies inside the continuous spectrum,
consistently with the results of \cite{RK1}. In Fig. \ref{fig2a} we
show the results obtained when viscosity is included in the
calculation.  In the Newtonian case (dotted line), the inviscid limit
is nicely recovered as before. As for the Cowling (dashed line) and RG
(solid line) calculations, we can follow the $r-$mode inside the
continuous spectrum until convergence to the inviscid limit is
reached.  The corresponding damping times are shown in
Fig. \ref{fig2b}. The Relativistic damping time lies between the
Newtonian and the Cowling calculation, and they agree within a factor
2.  For comparison we also include the estimate given by
Eq. ({\ref{tau0}}) using the average density of the star, which
overestimates the damping time of about 50 \%.  Notice that using the
average density in Eq. ({\ref{tau0}}), the damping time (for models
with constant $\eta$) depends only on $\rho R^2 \propto M/R$,
therefore it gives the same result for stars with the same
compactness.
  
\begin{figure}  
\includegraphics[height=.3\textheight]{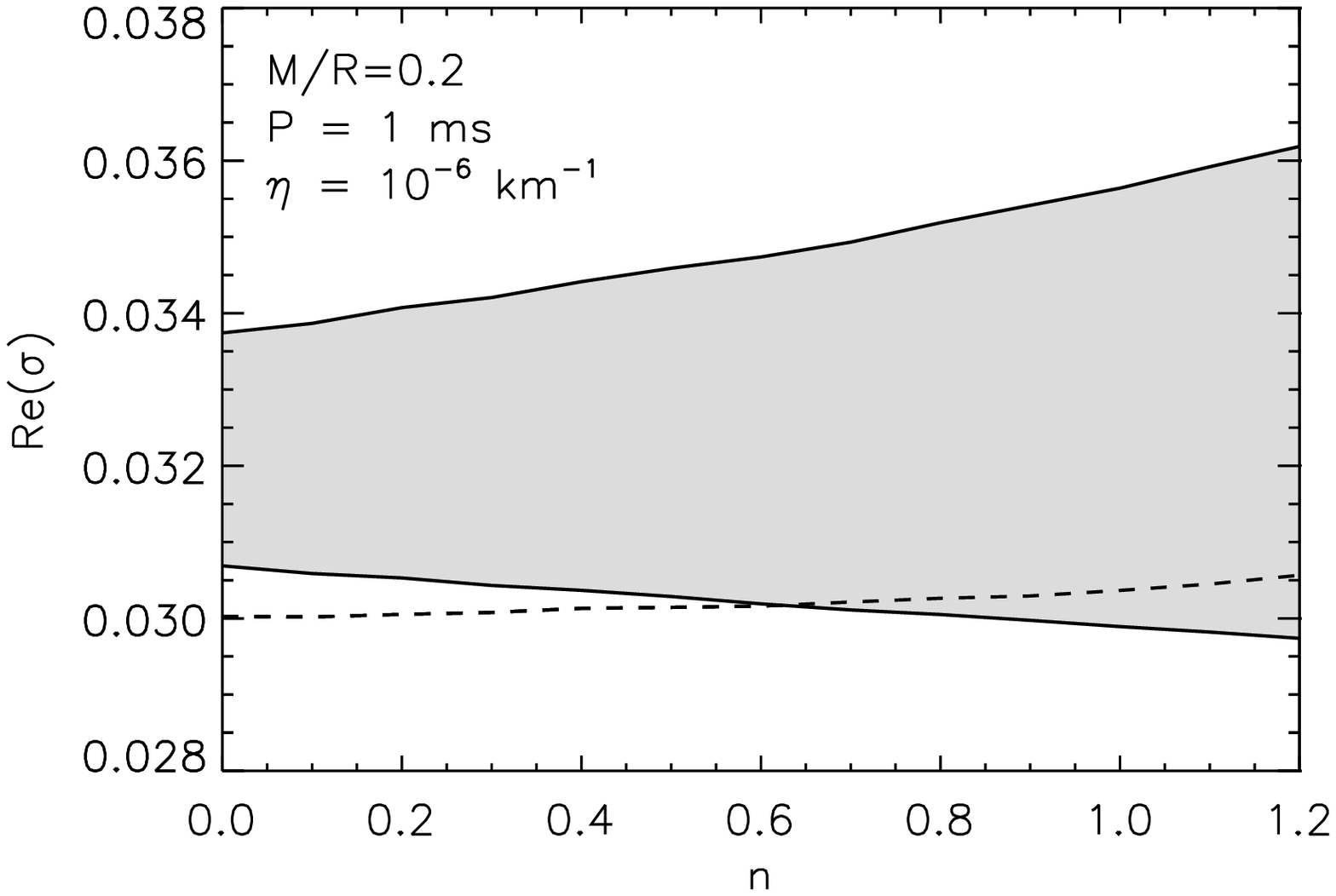}
\caption{Behaviour of the real part of the $r-$mode frequency  
($\sigma$) as a function of the polytropic index $n$ for models with a  
shear viscosity parameter $\eta=10^{-6}$ km$^{-1}$, compactness $M/R=0.2$ and  
period of 1 ms.  }  
\label{fig3}  
\end{figure}

In Fig. \ref{fig3} we show the behaviour of the 
real part of the $r-$mode frequency ($\sigma$) as a function of the 
polytropic index $n$ for models with compactness $M/R=0.2$, period of 
1 ms, and a shear viscosity $\eta=10^{-6}$ 
km$^{-1}$. The period and compactness have been chosen to allow for 
direct comparison with the results of \cite{RK1} who could not find 
the $r$-modes  for $n$ larger than a certain 
value, when the real part of the frequency reached the continuous 
spectrum. By introducing a small amount of viscosity, the frequency 
can be calculated for all polytropic indexes, or for any other stellar 
model, even if we stay at the most basic level of approximation: first 
order in the rotation parameter and neglecting the coupling between 
the axial and polar parts. 
Note that when the mode lays outside the continuous spectrum
we obtain results very similar  to \cite{RK1}.
\subsection{Realistic Neutron Stars}    
Here we consider stellar models constructed with realistic EOSs and
realistic viscosity profiles.  At low density (below $10^{12}$
g/cm$^3$) we use the BPS \cite{BPS} equation of state, while for the
inner crust $10^{12} <\rho < 10^{14}$ g/cm$^3$ we employ the SLy4 EOS
\cite{SLy4}.  At high density ($\rho > 10^{14}$ g/cm$^3$) we have
considered two different EOSs of neutron star matter representative of
the two different approaches commonly found in the literature:
potential models and relativistic field theoretical models. As a
potential model, we have chosen the EOS of
Akmal-Pandharipande-Ravenhall (\cite{APR}, hereafter APR). As an
example of the mean field solution to a relativistic Walecka--type
Lagrangian we have used the parametrization usually known as GM3
\cite{GM3}.

If the compactness is $M/R=0.2$, APR gives a mass of 1.53 $M_\odot$
and a radius of $11.3$ km, while GM3 gives $M=1.72 M_\odot$ and
$R=12.7$ km.  For a polytropic EOS with $n=1$ the corresponding mass
and radius for the same compactness is $M=1.74 ~ M_\odot$ and $R=12.8$
km.  For cold neutron stars below $10^9$ K, neutrons in the inner core
become superfluid and the dominant contribution to the shear viscosity
is electron-electron scattering (see e.g. the review \cite{AK}); in
this regime $\eta$ can be written as
\begin{eqnarray}
\eta_{ee}=6\times10^{18} \rho_{15}^2 T_9^{-2}~{\rm g/cm/s} = 
1.48\times10^{-15} \rho_{15}^2 T_9^{-2}~{\rm km}^{-1}
\label{etaee}
\end{eqnarray}
with $\rho_{15}$ and $T_9$ being the density and temperature in units
of $10^{15}$ g/cm$^3$ and $10^{9}$ K.  Having this in mind, we have
used a viscosity coefficient with a quadratic dependence on density
\begin{eqnarray}
\eta= \eta_0 \left(\frac{\rho}{\rho_0}\right)^2 ~{\rm km}^{-1} ~,
\label{eta0}
\end{eqnarray}
where $\rho_0$ is the central density. Since old neutrons stars are
nearly isothermal, we have parametrized our results as a function of
the constant $\eta_0$ that includes the temperature dependence.

\begin{figure}  
\includegraphics[height=.3\textheight]{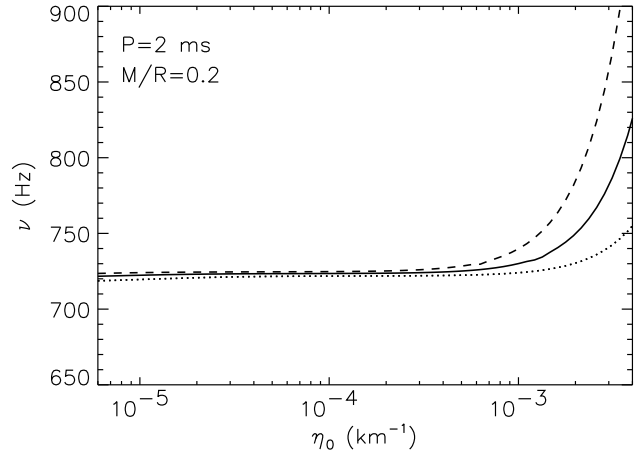}
\caption{Comparison of the $r-$mode frequency as a function of the  
the viscosity coefficient at the center, $\eta_0$, for a polytropic star with $n=1$  
(dotted line), the APR (dashed line) and the GM3 (solid line)  
equations of state. The compactness parameter in all cases is  
$M/R=0.2$ and the period is 2 ms.  }  
\label{fig4a}  
\end{figure}

In Fig. \ref{fig4a} we show the $r-$mode frequency as a function of
$\eta_0$, comparing the three EOSs: polytrope (dots), APR (dashes) and
GM3 (solid lines). In all cases the rotation period is $P=2$ ms. As we
can see in the figure, the frequency is rather insensitive to the
particular details of the EOS, provided that the rotation frequency
and $M/R$ are the same.  The Newtonian limit depends only on the
angular frequency ($\Omega$) and the relativistic correction enters
through the frame dragging. Since this correction goes as
$\omega/\Omega \approx I/R^3 \propto M/R$, the leading order
contribution to the frequency is again a function only of the
compactness.

\begin{figure}  
\includegraphics[height=.3\textheight]{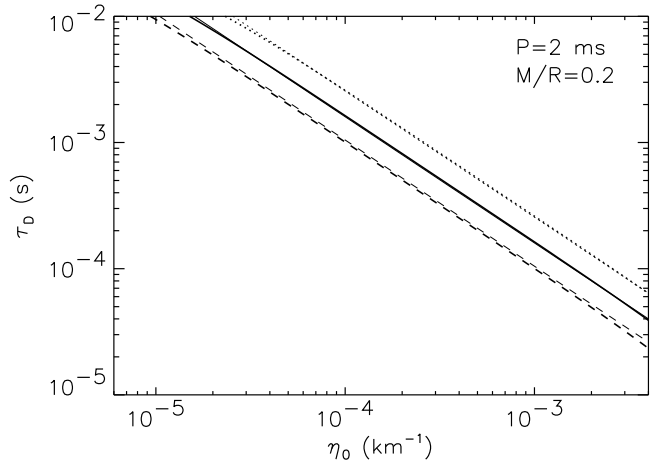}
\caption{Comparison of the $r-$mode damping time as a function of 
$\eta_0$ for a $n=1$ polytrope 
(dotted line), the APR (dashed line) and the GM3 (solid line)  
equations of state. The compactness parameter in all cases is  
$M/R=0.2$ and the period is 2 ms.  The thin lines show the estimates
given by Eq. (\ref{taugood}), which are difficult to distinguish from
the real results. }  
\label{fig4b}  
\end{figure}

In Fig. \ref{fig4b} we show the damping times for the same models as
in Fig. \ref{fig4a}.  For constant density stars, the $\approx \rho
R^2$ dependence of the damping time translates into a $M/R$ dependence
for models with constant $\eta$. In realistic, cold ($T\lesssim
10^9$K) neutron stars, the density is not constant and the viscosity
is dominated by the electron-electron scattering process
(\ref{etaee}). Thus, the analytic result (\ref{tau0}) underestimates
the viscous damping time. An improved calculation for $n=1$ polytropes
taking into account the density profiles (Andersson \& Kokkotas 2001)
gives
\begin{eqnarray}
\tau = 2.2 \times10^{7} \left(\frac{1.4 M_{\odot}}{M}\right)
\left(\frac{R}{10\,{\rm km}}\right)^5 T_9^2~{\rm s}.
\end{eqnarray}
Assuming a general dependence of the viscosity of the form of Eq. (\ref{eta0}),
the above damping time can be shown to satisfy
\begin{eqnarray}
\tau = 3.26 \times10^{-8} \left(\frac{1.4 M_{\odot}}{M}\right)
\left(\frac{R}{10\,{\rm km}}\right)^5 
\left(\frac{\rho_0}{10^{15}{\rm g/cm}^3}\right) \frac{1}{\eta_0} ~{\rm s}.
\label{tauak}
\end{eqnarray}
Our relativistic calculations show approximately the same dependence
on mass and radius as Eq. (\ref{tauak}), but the damping times are
systematically larger of about 60\%.  We found that a better fit for
the realistic neutron stars (APR, GM3) as well as for the $n=1$
polytrope is given by
\begin{eqnarray}
\tau = 5.22 \times10^{-8} \left(\frac{1.4 M_{\odot}}{M}\right)
\left(\frac{R}{10\,{\rm km}}\right)^5 
\left(\frac{\rho_0}{10^{15}{\rm g/cm}^3}\right) \frac{1}{\eta_0} ~{\rm s}.
\label{taugood}
\end{eqnarray}
In Fig. \ref{fig4b} we show, together with the numerical results
(thick lines), the results corresponding to the previous fit (thin
lines). The good agreement between them is apparent.
\begin{theacknowledgments}
We thank Adamantios Stavridis, Kostas Kokkotas and Nils Andersson for 
useful comments and discussions. This work has been supported by the 
Spanish MEC grant AYA-2004-08067-C03-02, and the {\it Acci\'on 
Integrada Hispano--Italiana} HI2003-0284.  J.A.P.~is supported by a 
{\it Ram\'on y Cajal} contract from the Spanish MEC. 
\end{theacknowledgments}

   
\end{document}